\documentclass[useAMS,usenatbib]{lib/mnras}

\usepackage{natbib}
\usepackage{epsfig}
\usepackage{scrtime}
\usepackage{graphicx}	
\usepackage{amsmath}	
\usepackage{amssymb}	
\usepackage{float}
\usepackage[caption = false]{subfig}
\usepackage{multirow}
\usepackage{rotating}
\usepackage{ulem}
\usepackage{booktabs}
\usepackage{caption} 
\usepackage{threeparttable} 
\usepackage{threeparttablex}
\usepackage{fancyhdr} 
\usepackage{longtable}
\usepackage{lscape}
\usepackage{xcolor}
\usepackage{ulem}
\usepackage{multibib}
\usepackage{soul}

\newcommand{\mincir}{\raise-3.truept\hbox{\rlap{\hbox{$\sim$}}\raise4.truept\hbox{$<$}\ }}

\title[Youngest Star Clusters in the LMC]{The Youngest Star Clusters in the Large Magellanic Cloud}
\author[Ch\'avez et al.] {R.\, Ch\'avez$^{1, 2}$\thanks{E-mail: r.chavez@irya.unam.mx}, R.\ A.\, Gonz\'alez-L\'opezlira$^{1}$ and G.\, Bruzual$^{1}$ \\
$^{1}$Universidad Nacional Autónoma de México, Instituto de Radioastronomía y Astrofísica, 58090, Morelia, Michoacán, México \\
$^{2}$Secretaría de Ciencia, Humanidades, Tecnología e Innovación, Av. Insurgentes Sur 1582, 03940, Ciudad de México, México \\
}

\begin{document}
\date{MN-24-0876-MJ.R2 --- Compiled at \thistime\ hrs  on \today\ }
\maketitle
\label{firstpage}
\begin{abstract}
This study presents a comprehensive analysis of the youngest stellar clusters in the
Large Magellanic Cloud (LMC), utilising a multi-wavelength approach. We analyse
data spanning from infrared to ultraviolet wavelengths, with the goal of enhancing our understanding of
these clusters’ physical properties, such as age, mass, and size. Our methodology
includes a novel cluster detection procedure; it employs machine learning techniques 
for the accurate identification of these young clusters. The Markov Chain Monte Carlo analysis, using the Automated Stellar Cluster Analysis tool, plays a crucial role in deriving
the clusters’ key physical parameters. Our findings provide significant insights into
the early stages of stellar and galactic evolution, particularly in dwarf galaxy environments, and contribute to the broader understanding of star formation and cluster
evolution. For the 109 clusters in our sample younger than 5 Myr, we measure a positive correlation between cluster mass and the mass of the most massive star in the cluster. This study emphasises the importance of multi-wavelength observations in revealing the intricate properties of young stellar clusters.
\end{abstract}


\section{Introduction}
Massive young stellar clusters (YSCs) represent crucial environments for investigating the complexities of star formation (SF), and its interplay with the interstellar medium (ISM). These clusters are not merely stellar cradles, but also epicentres of significant astrophysical phenomena. Supernovae (SN) and stellar winds within YSCs exert a profound influence on their surroundings. These events are instrumental in catalysing shock waves, disintegrating dust grains and molecules, and compressing molecular clouds, thereby playing a critical role in fostering subsequent star formation \citep{2000ASPC..211...25C, Schneider_2020, 2010ApJ...709..191M, 2012MNRAS.421.3488H}.

Furthermore, YSCs are integral to the chemical evolution of galaxies. They facilitate the dispersal of newly synthesised elements into the ISM, thereby modifying its chemical composition. Hence, the study of YSCs offers a unique vantage point for examining stellar evolution and the life-cycle of galaxies \citep{2023MNRAS.526.1713S, 2023MNRAS.522.3092L, 2023ApJ...948...65L}.

Last but not least, it is in YSCs where one can measure the stellar initial mass function (IMF) before it is modified by stellar evolution. It is there where it is possible to discern whether the IMF is a scale-invariant probability distribution function from which stellar masses are stochastically sampled  or, rather, that stellar masses are related to the conditions in which SF happens \citep[e.g.,][and references therein] {2023A&A...670A.151Y}.

The study of YSCs, thus, provides a window into the early stages of stellar and galactic development. By examining these clusters, we can gain a deeper understanding of the processes that shape galaxies and the universe at large. The dynamic interplay of SF, SN, and stellar winds within YSCs offers a rich tapestry of astrophysical processes, making these clusters invaluable environments for astrophysical research \citep{2017MNRAS.466.1903G, 2021AJ....162..236M, 2023MNRAS.522.3092L}.

The Magellanic Clouds (MC) harbor the nearest examples of massive
YSC, and of SF under this regime. Given their well constrained distances, i.e., for the LMC a distance modulus \sout{of} $\mu = 18.476 \pm 0.002$ mag \citep{2019Natur.567..200P} and for the SMC $\mu = 18.977 \pm 0.032$ mag \citep{2020ApJ...904...13G}, and the fact that we can observe them with exceptional spatial and spectral resolutions, they are excellent laboratories to understand with great detail the effects of massive YSC on the ISM, and to gather information about the stellar and cluster initial mass functions  \citep{Bitsakis2017}. The Large Magellanic Cloud (LMC) hosts the well known giant extragalactic HII 
region (GEHR) 30 Doradus (the Tarantula Nebula), the largest and most massive HII region complex in the Local Group (LG); the nebula is $\sim$15$'$ (200 pc) in diameter. Meanwhile, in the Small Magellanic Cloud (SMC) the largest GEHR is N66/NGC 346.

The LMC, as one of the most prominent and closest members of the Local Group, has been the subject of numerous multi-wavelength studies, ranging from the optical and ultraviolet (UV) to the infrared (IR) wavelengths. These works have contributed to a comprehensive understanding of the star formation processes, the stellar populations, and the ISM properties in the LMC.

UV observations of the LMC have revealed complex details about its extinction properties. A reanalysis of the LMC's UV extinction using data from the International Ultraviolet Explorer \citep[IUE;][]{bogg78a, bogg78b} satellite final archive has shown significant differences in the average extinction curves of various regions within the LMC, particularly in the strength of the 2175 \AA{} bump \citep{Misselt_1999}.

IR studies, particularly with the Spitzer Space Telescope \citep{wern04}, have been instrumental in identifying young stellar objects (YSOs) in the LMC. A thorough analysis using Spitzer's Infrared Array Camera \citep[IRAC;][]{Fazio2004} and Multiband Imaging Photometer \citep[MIPS;][]{Rieke2004} has identified a sample of potential YSOs in the LMC, shedding light on the star-forming ISM's evolution across cosmic time \citep{Gruendl_2009}.

Optical spectroscopic analyses of Wolf-Rayet (WR) stars in the LMC have provided insights into the properties and evolution of massive stars. An exhaustive analysis of the WN class of WR stars in the LMC has revealed the presence of two distinct groups, based on their luminosities and hydrogen content \citep{2014A&A...565A..27H}.

In the present in-depth study, we delve into the exploration of the youngest stellar clusters within the LMC. Our primary objective is to elucidate their physical properties, examine the underlying scaling laws, and analyse their IMF. This research contributes significantly to our understanding of stellar evolution and cluster dynamics in a prototypical dwarf galaxy environment. 

\section{Data Samples}
In order to identify the youngest stellar clusters in the LMC and gain a good comprehension of their physical properties, we use a wide range of photometric archival data, from the radio to the UV. In the following subsections we briefly describe the origin and characteristics of the data. 

\subsection{Infrared} 
We use the IR data from the project \textit{Surveying the Agents of a Galaxy’s Evolution} \citep[SAGE;][]{Meixner2006}, which consisted in a uniform and unbiased imaging survey of the LMC using the Spitzer Space Telescope's IRAC and MIPS instruments. This extensive survey aimed to study the ISM and stars in the LMC, and to provide insights into the life cycle of dust in this unique environment. The SAGE survey produced mosaics with IRAC at the wavelengths of 3.6, 4.5, 5.8, and 8.0 \(\mu\)m, and with MIPS at 24, 70, and 160 \(\mu\)m. The survey's observations have enabled a detailed understanding of dust processes in the ISM, the current star formation rate in the LMC, and the rate at which evolved stars inject mass into the LMC's ISM. 

Likewise, the similar legacy program Surveying the Agents of Galaxy Evolution in the Tidally Stripped, Low Metallicity Small Magellanic Cloud \citep[SAGE-SMC;][]{2011AJ....142..102G} has been instrumental in identifying a significant population of YSOs in the SMC, and its relation to gas tracers \citep{sewi13}.

\subsection{Optical}
The optical data for this study are taken from the Survey of the Magellanic Stellar History (SMASH) DR2 \citep{Nidever2017}. Observations for this survey were obtained with the Dark Energy Camera \citep[DECam;][] {2015AJ....150..150F}, which is mounted at the prime focus of the Blanco 4-m telescope at the Cerro-Tololo International Observatory (CTIO). DECam is equipped with a large, focal plane, mosaic CCD and a five-element optical corrector. The survey covers an area of 480 deg$^2$ in the sky, and reaches depths in $ugriz$ of $\sim$ 24 mag. Additionally, the data from this survey have an astrometric precision of approximately 20 mas.

\subsection{UV}
In this study, we exploit the  near-UV (NUV) band data presented by \citet{Simons2014}. This dataset is an aggregation of archival observations secured by the Galaxy Evolution Explorer \citep[GALEX;][]{Martin2005}. The mosaic compiled from these observations encompasses an area of \(15\ \mathrm{deg^2}\) within the  LMC. Notably, this coverage omits the bar region of the LMC, a significant structural component of the galaxy. To make up for this, observations of the bar area were sourced from the Swift Ultraviolet-Optical Telescope \citep[UVOT;][]{romi05} Magellanic Clouds Survey, commonly referred to as SUMAC \citep{Siegel2014}. The combination of data from both GALEX and SUMAC provides a comprehensive view of the LMC in the NUV band, facilitating a detailed analysis of its stellar populations and structural components.

\section{Cluster Detection Methodology}
We introduce an innovative methodology tailored for the identification of the youngest star clusters within the LMC. This approach hinges on a simultaneous multi-wavelength analysis, encompassing NUV, optical band measurements in the {\it ugriz} bands, and IR photometry at 8 and 24 $\mu$m. 

The selection of these specific bands follows a strategy devised to capture the distinct phenomenology associated with the youngest star clusters. Observations in the 24 $\mu$m band are especially valuable, as they allow for the detection of clusters still  enshrouded in their natal gas clouds \citep{2014ApJ...789...81F}. The NUV band, on the other hand, is instrumental in pinpointing the presence of young, massive, blue OB stars, which are indicative of recent star formation activity \citep{2022MNRAS.517.1518J}. Additionally, the $ugriz$ optical bands serve a dual purpose: they not only aid in the identification process, but also facilitate a deeper understanding of the intrinsic physical properties of the star clusters. For example, ages, metallicities, and extinctions of individual clusters are measured through isochrone-fitting of optical colour-magnitude diagrams (CMD; see Section 4).

\begin{figure*}
\begin{center}
\includegraphics[width=2.0\columnwidth]{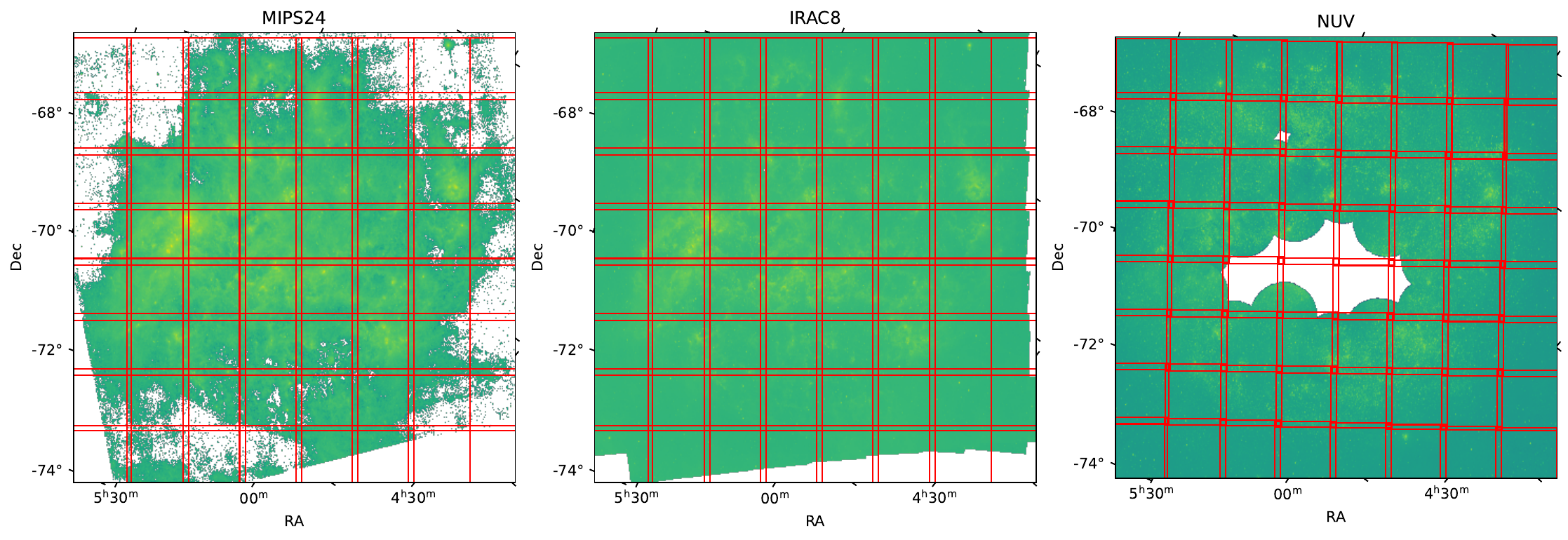}
\end{center}
        \caption{8 $\times$ 8 deg$^2$ LMC mosaic in three bands. {\it Left-hand panel:} MIPS (Spitzer) 24 $\mu$m. {\it Central panel:} IRAC (Spitzer) 8 $\mu$m. {\it Right-hand panel:} GALEX NUV. In red we show a grid of sub-regions of 66 $\times$ 66 arcmin$^2$, which were analysed individually. The colour map represents the flux density in each pixel; yellow regions are brighter.}
\label{fig:C02}
\end{figure*}

 To enhance the efficiency of the YSC identification, we subdivided the entire analysis region into subregions, each spanning \(66 \times 66\) arcmin$^2$. This subdivision was designed to ensure overlaps at the boundaries of adjacent subregions, mitigating potential information loss. This approach enabled the parallelisation of our algorithm, allowing for concurrent analysis of different subregions across multiple computer cores. For this parallelisation, we utilised the Python library \textit{joblib}. Figure \ref{fig:C02} depicts the grid of these subregions, overlaid on the original \(8 \times 8\) deg mosaic, indicating the individual areas subjected to analysis.

\begin{figure*}
\begin{center}
	\includegraphics[width=2\columnwidth]{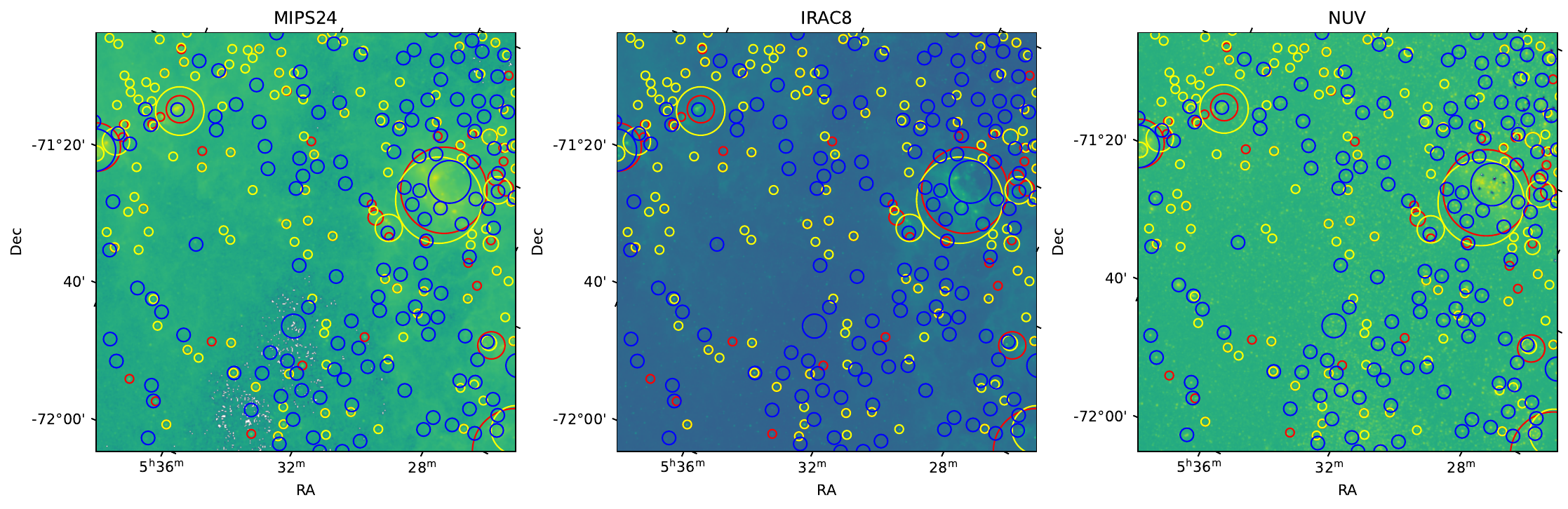}
\end{center}
	\caption{One of the typical individual sub-regions analysed from the LMC, in three bands. {\it Left-hand panel:} MIPS (Spitzer) 24 $\mu$m. {\it Central panel:} IRAC (Spitzer) 8 $\mu$m. {\it Right-hand panel:} GALEX NUV. In the three panels we show the same 66 $\times$ 66 arcmin$^2$ analysed region. In all the panels, red circles mark detections made in the 24 $\mu$m band, yellow circles detections in the 8 $\mu$m band, blue circles detections in the NUV band. Larger circles indicate larger detected cluster size}. 
\label{fig:C03}
\end{figure*} 

For the identification of sources within each analysed subregion of our infrared (8 and 24 µm) and near UV astronomical images, we employ a two-step process involving Gaussian normalisation, followed by a Laplacian of Gaussian (LoG). This technique is particularly beneficial in astronomical imaging for accentuating specific features, notably star clusters in this context. Initially, the image is convolved with a Gaussian kernel, a process that smooths the image to reduce noise and minimise the detection of spurious details. Following this, the Laplacian operator is applied to the smoothed image. The Laplacian, a second derivative operator, enhances regions of rapid intensity change, effectively highlighting the edges of stellar clusters. This method significantly aids in the precise delineation and identification of stellar clusters within the subregions \citep{marr1980theory, kawalec2014edge}.

We have implemented this method using the \textit{scikit-image} algorithm suite, a Python-based library known for its sweeping image processing capabilities \citep{scikit-image}. In order to accommodate the diverse sizes of star clusters, we utilise Gaussian kernels with varying values of standard deviation ($\sigma$). This approach allows us to detect clusters of different sizes by adjusting the $\sigma$ of the Gaussian kernel, which controls the extent of smoothing and, consequently, the sensitivity to clusters of various scales. The choice of $\sigma$ values is tailored to each band's angular resolution in the infrared (8 and 24 µm) and NUV, ensuring that the smallest detectable cluster size is optimised based on the inherent resolution limits of the data. This methodology not only enhances our ability to detect clusters across a spectrum of sizes, but also guarantees that our detection thresholds are  calibrated to the observational constraints of each specific band.

Figure \ref{fig:C03} offers a visual representation of our approach, highlighting our detections across three specific bands, i.e., 24 $\mu$m, 8 $\mu$m, and NUV, centred on a typical region.

We should notice that we do not expect to find exactly the same YSCs than previous studies, because in general we use different bands for the analysis. As an example, \citet{Bitsakis2017} employed NUV and 3.6$\mu$m (near-IR) data, thus selecting, in general, older clusters than ours and with a larger age dispersion.

In the subsequent phase of our analysis, after obtaining detections across all bands and assimilating catalogue data, we engage a sophisticated clustering approach to refine our results. Specifically, we utilise the Balanced Iterative Reducing and Clustering using Hierarchies (BIRCH) algorithm \citep{Zhang1996} to conduct a hierarchical clustering analysis. This method is particularly advantageous for large datasets, as it incrementally and dynamically clusters incoming multidimensional metric data points to produce a dendrogram, i.e., a tree-like structure. 

The primary objective of this clustering is to ascertain the reliability of our detections. By analysing the overlap and coincidence of detections across different bands and with existing catalogues, we can assign a probability score to each detection. This score quantifies the confidence level of a particular detection, providing a robust metric to differentiate between potential true detections and spurious ones.

For each detection in our study, we obtain photometry through a circular aperture. This is a widely-used technique to measure the flux within a circular region centred on the source, while accounting for the background noise \citep{howell2006handbook}. To implement this, we use the \textit{Photutils} package, an integral component of the \textit{astropy} library \citep{Bradley2019, price2018astropy}.

The result of this meticulous process is the creation of a highly refined catalogue of clusters, with 3962 objects, detailed in Appendix A. For each cluster, the catalogue includes critical information, such as the centroid coordinates, which accurately determine the cluster's location, along with its physical dimensions, core and tidal radii, age, internal extinction, and mass.

This catalogue serves as a foundational resource for further investigations, compiling data that paves the way for deeper analysis. The physical parameters of the clusters, derived from photometric data (as outlined in the following section), render this catalogue an invaluable tool for advancing our understanding of stellar cluster characteristics, and their roles within the larger frameworks of star formation and galaxy evolution \citep{lada2003embedded}.

\section{Cluster Physical Properties}

\begin{figure*}
\begin{center}\includegraphics[width=2.0\columnwidth]{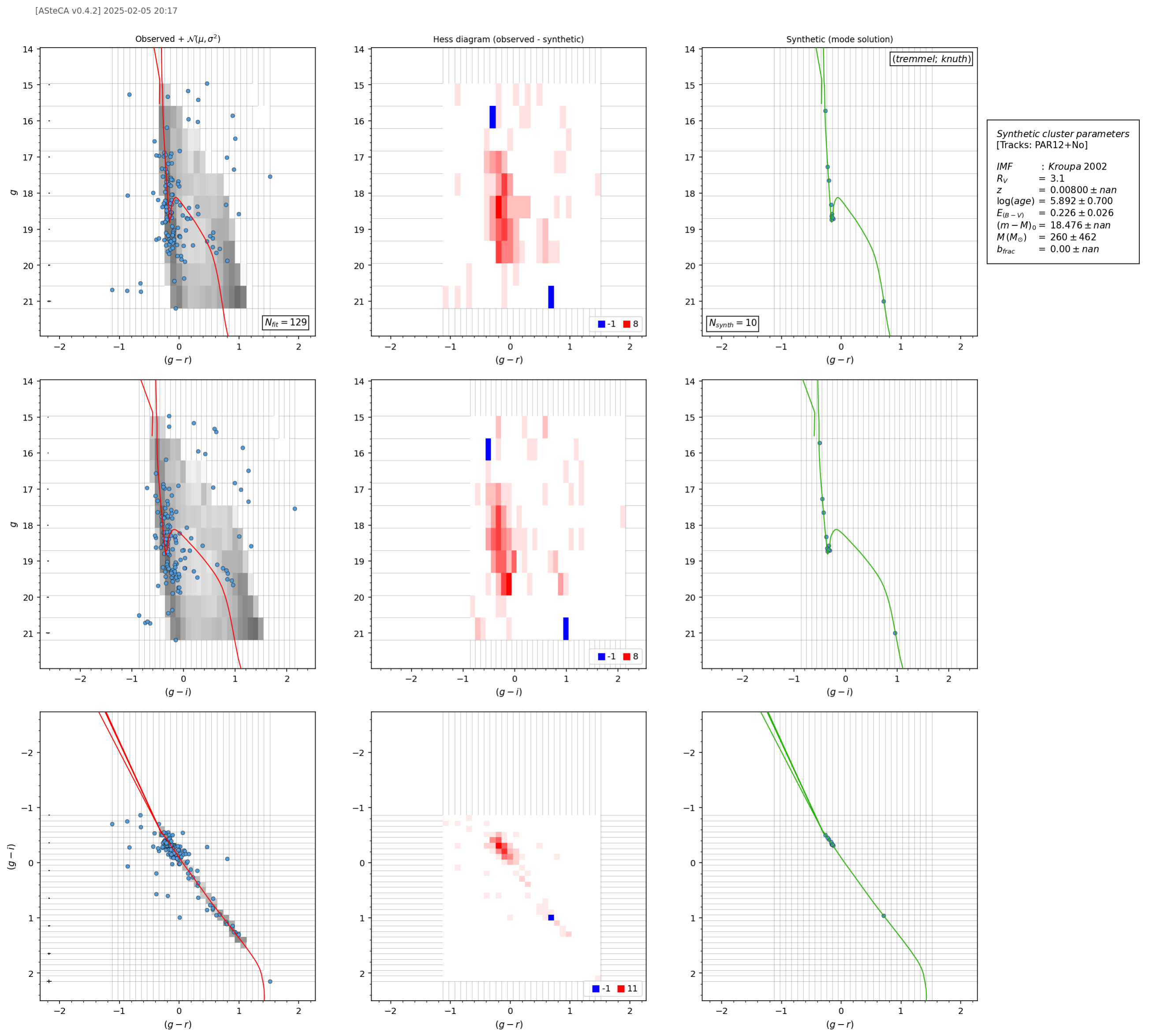}
\end{center}
	\caption{Physical parameters of cluster J053655.49-701608.40. {\it Upper left:} observed \(g-(g-r)\) CMD; {\it blue dots:} observed stars; {\it red line:}  best-fitting isochrone; {\it grey shaded background:} \(1\sigma\) deviation around the isochrone, factoring-in binning in colour and magnitude, and star counts per bin ($\mathcal N$ ($\mu$, $\sigma^2$)); {\it inset:} star counts used for fitting. {\it Upper right panel:} mirrors the left upper panel, but for a synthetic cluster derived from the best-fitting isochrone, following a Poisson likelihood ratio matching to the observed cluster \citep{tremmel2013modeling}; {\it blue dots:} synthetic cluster members; {\it inset:} member count. The adjacent frame details the synthetic cluster's physical properties. {\it Upper central panel:} Hess diagram contrasting the stellar density between observed and synthetic clusters. {\it Middle panels:} like upper ones, for the \(g-(g-i)\) CMD. {\it Bottom panels:} like upper and middle panels, for the \((g-i)-(g-r)\) colour-colour diagram.
}
\label{fig:params}
\end{figure*}

Once we identify the stellar clusters, through the techniques described in the previous section, on images from the infrared and NUV bands, we proceed to refine our analysis with optical data from the SMASH survey on the $gri$ bands. Specifically, we use the centroids and radii of the identified clusters to locate individual stars within a radius expanded by $\sqrt{3}$ times the identified cluster radius. This method effectively captures all potential cluster members by encompassing a broader area around the core of each cluster, thereby ensuring a comprehensive selection of stars within the cluster's influence. This approach leverages the high resolution and broad wavelength coverage of the SMASH data to accurately assess the stellar populations and characteristics of these clusters.

Once we identify the stellar clusters at the LMC, we embark on a Markov Chain Monte Carlo (MCMC) analysis of the data. This analysis focuses on the optical \(g\) magnitude, as well as on the (\(g-r\)) and (\(g-i\)) optical colours of individual sources within each cluster. Our primary objective is to derive and constrain various parameters of the cluster, including its size, age, extinction, and mass. For this purpose, we employ the Automated Stellar Cluster Analysis \citep[ASteCA;][]{Perren2015}, a comprehensive suite of tools designed to automate the analysis of stellar clusters. ASteCA performs a range of standard tests to determine the basic parameters of stellar clusters, including centre coordinates, radius, luminosity function, and integrated colour magnitude. Utilising positional data, the algorithm accurately determines the cluster's centre using a two-dimensional Gaussian kernel density estimator, and calculates the radius through a radial density profile. ASteCA also incorporates a Bayesian field star decontamination method that assigns membership probabilities based solely on photometric data. The Bayesian analysis in ASteCA utilises the emcee sampler \citep{2013PASP..125..306F}, a powerful and efficient MCMC ensemble sampler. Furthermore, it includes an isochrone fitting procedure that generates synthetic clusters from theoretical isochrones. This process, which employs a genetic algorithm to select the best fit, allows the estimation of the cluster's metallicity, age, extinction, mass and distance, along with their uncertainties. In this work we fit simultaneously age, extinction and mass, leaving fixed the values of distance, $\mu = 18.476$, and metallicity, $Z = 0.008$ \citep[cf.][]
{swb80,cole2005spectroscopy}. 

ASteCA also features tools for statistically estimating the probability that a cluster is a true physical cluster rather than a random overdensity of field stars. The algorithm has been validated on a large set of synthetic clusters and a smaller set of observed Milky Way open clusters, demonstrating its ability to recover cluster parameters with acceptable precision even in the presence of significant field star contamination.

ASteCA addresses the issue of field star contamination in crowded regions using a Bayesian decontamination algorithm (DA). This method assigns membership probabilities (MPs) to stars by comparing the observed cluster region with the surrounding field regions. Unlike traditional approaches based on proximity to the main sequence or proper motion data, ASteCA estimates MPs using photometric data alone, making it applicable even in highly contaminated areas.

The Bayesian DA defines three regions: the observed cluster region ($C$), a nearby field region ($B$), and an idealised clean cluster region ($A$), such that $A + B = C$. Using the Bayes theorem, the probability that a star belongs to the cluster is computed as:
    \begin{equation}
    MP_j = \frac{L_B,j \cdot (N_A / N_B)}{(N_A / N_B) L_A,j + L_B,j},
    \end{equation}
where $L_A$ and $L_B$ are likelihood functions based on the star’s magnitude and colour distributions. The process iterates over multiple field regions, refining the membership probabilities by averaging between different realisations of $A$. This statistical approach allows robust decontamination even when the density of the field star varies significantly in the observed frame.

The theoretical isochrones employed for the analysis are from the PARSEC library version 1.2S \citep{2014MNRAS.445.4287T, 2014MNRAS.444.2525C}. These isochrones are sampled in a metallicity range of $0.001 \leq Z \leq 0.012$, in steps of $\delta Z = 0.0002$, and an age interval $5.0 \leq \log(\mathrm{age}) \leq 10.4$, in steps of $\delta \log(\mathrm{age}) = 0.02$.

To estimate the cluster's mass, ASteCA uses an IMF to generate a synthetic population of stars that matches the observed CMD, in this case the IMF by \citet{2002Sci...295...82K}. The total mass of the synthetic cluster is set to a default value, ensuring that all stages of stellar evolution are represented. The genetic algorithm then adjusts the mass along with other parameters to best fit the observed data. The result is an estimate of the cluster's mass that, along with the other parameters, provides a complete characterization of the cluster.

Figure \ref{fig:params} provides a representative illustration of the constraints procured, using ASteCA, on the physical attributes of a specific LMC YSC, as identified through our methodology. The upper left panel of this figure displays the observed \(g-(g-r)\) CMD for the cluster J053655.49-701608.40. Here, the blue dots represent observed stars, while the red line delineates the isochrone that best fits the data. The surrounding grey shaded background indicates the \(1\sigma\) deviation around the isochrone, taking into account the binnings in colour and magnitude, and the number of stars in each bin ($\mathcal N$ ($\mu$, $\sigma^2$)). An inset reveals the number of stars used for the fitting. In all the panels of this figure, binning is executed automatically, adhering to the Knuth optimal data-based binning \citep{2006physics...5197K}.

The upper right panel similarly portrays the CMD, but for a synthetic cluster derived from the best-fitting isochrone. Blue dots, in this context, denote members of this synthetic cluster, which align perfectly with the isochrone. The inset displays the member count of the synthetic cluster. The matching between synthetic and observed clusters is achieved through a Poisson likelihood ratio, as defined by \citet{tremmel2013modeling}. In a frame adjacent to this panel, the physical properties of the synthetic cluster are enumerated.

The upper central panel introduces a Hess diagram, highlighting the stellar density difference in each bin between observed and synthetic clusters.
The three middle panels echo the upper panels' information, but for the \(g-(g-i)\) CMD. In contrast, the three bottom panels convey analogous data for the \((g-i)-(g-r)\) colour-colour diagram (CCD).

\begin{figure}
\begin{center}
        \includegraphics[width=1.0\columnwidth]{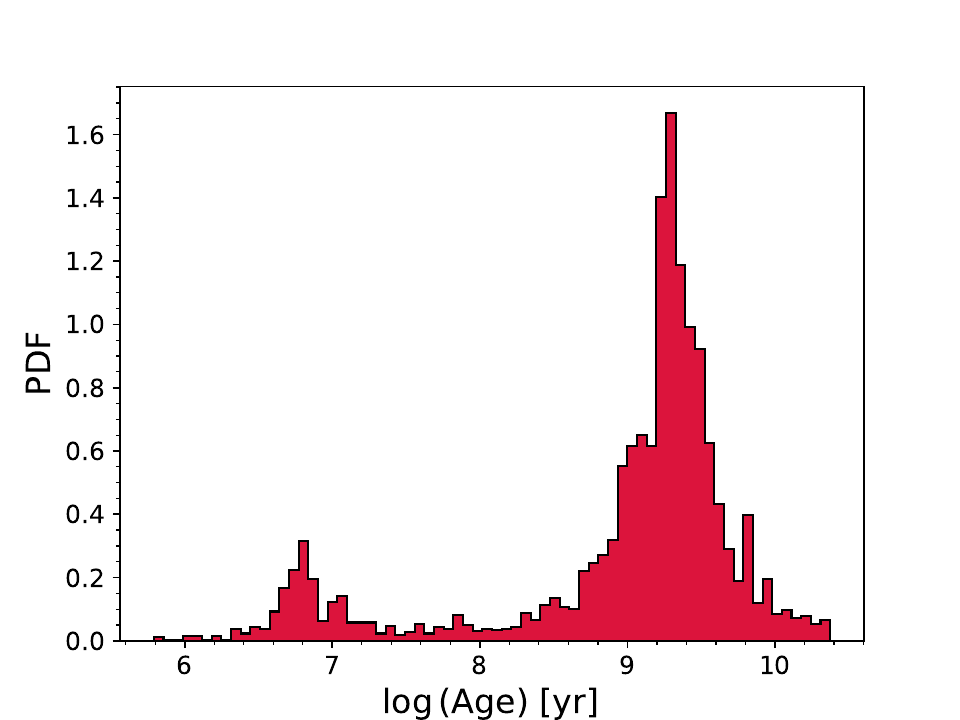}
\end{center}
	\caption{Distribution of star cluster ages in our sample.}
\label{fig:Hage}
\end{figure}

In Figure \ref{fig:Hage}, the age distribution of our cluster sample is presented. A distinct bimodality is evident. The majority of the objects exhibit an age range between \(10^9\) and \(10^{10}\) years. However, a secondary, younger component with ages of \(10^7\) years or less is also discernibly present.

\begin{figure}
\begin{center}
        \includegraphics[width=1.0\columnwidth]{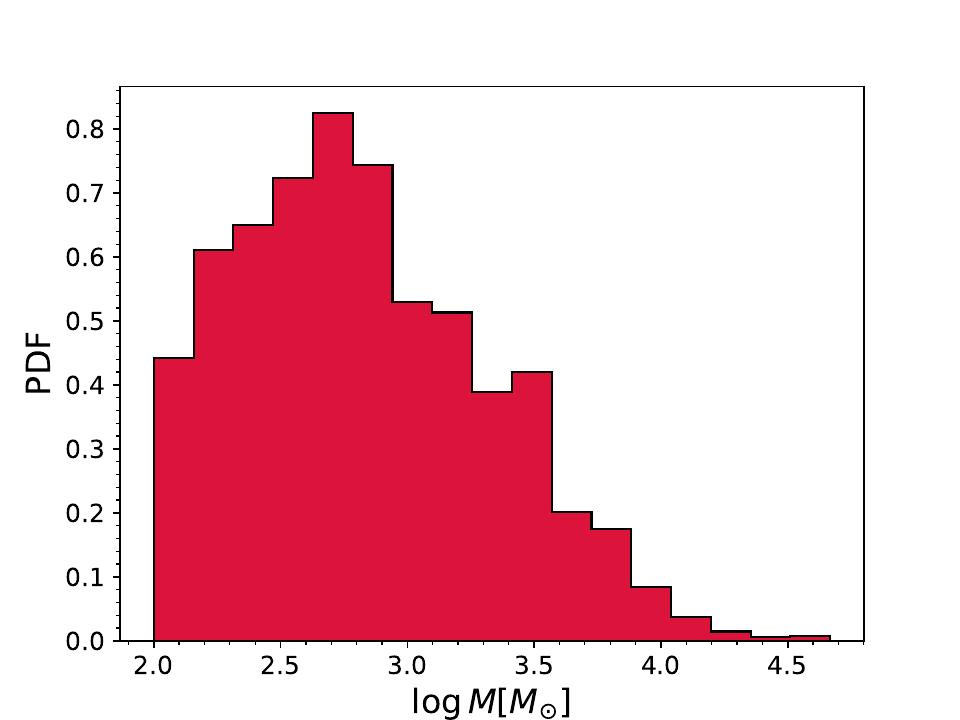}
\end{center}
	\caption{Distribution of star cluster masses in our sample.}
\label{fig:HMass}
\end{figure}

In Figure \ref{fig:HMass}, we depict the mass distribution of our star cluster sample. The distribution displays a pronounced peak at approximately \(10^3\ M_{\odot}\). Notably, a majority of the clusters reside below this mass, while there exists a marked tail extending towards higher masses.

\begin{figure}
\begin{center}
        \includegraphics[width=1.0\columnwidth]{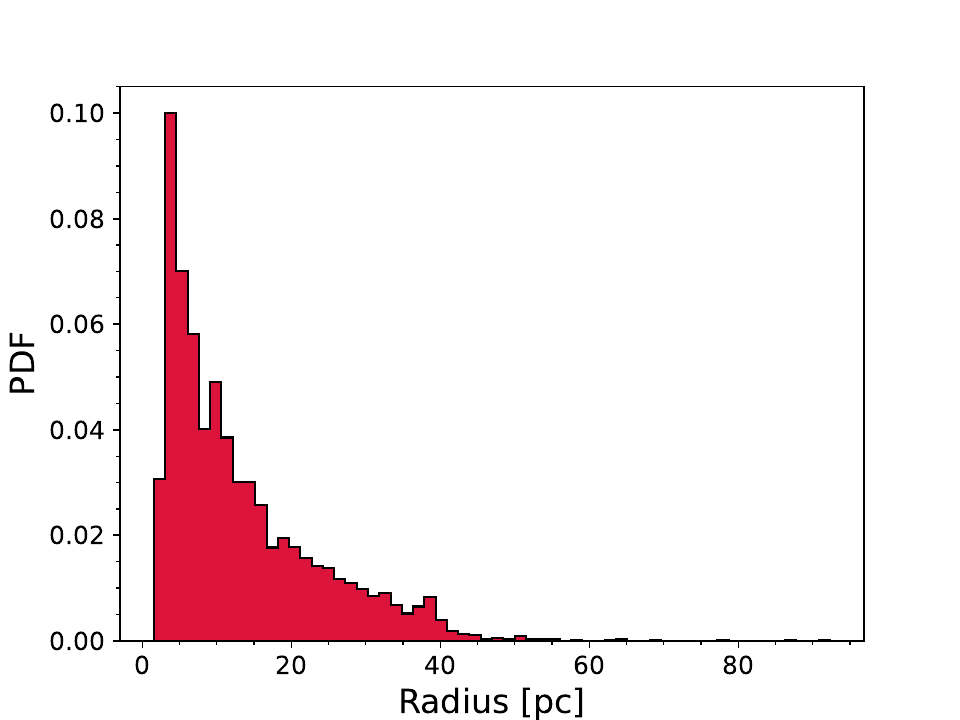}
\end{center}
	\caption{Distribution of star cluster radii in our sample.}
\label{fig:Hrad}
\end{figure}

In Figure \ref{fig:Hrad}, the distribution of radii of our star cluster sample is illustrated. The distribution prominently peaks at approximately 8 pc. While there is an extended tail reaching radii of about 40 pc, a sparse number of objects sport radii of up to \(\sim 86\) pc.

\begin{figure}
\begin{center}
        \includegraphics[width=1.0\columnwidth]{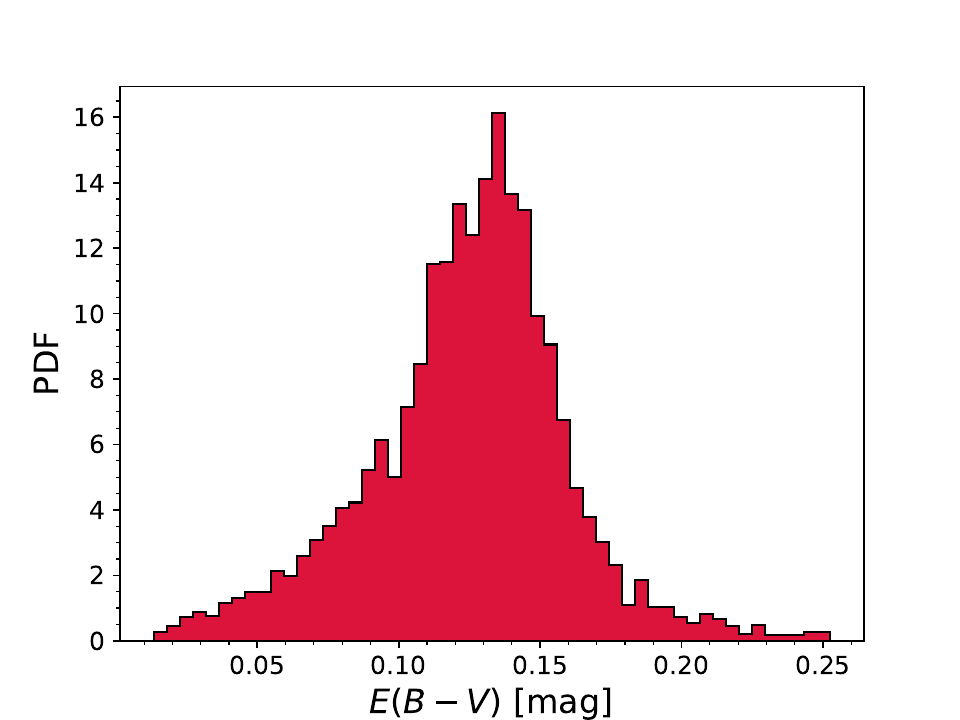}
\end{center}
	\caption{Distribution of star cluster reddening in our sample.}
\label{fig:HExt}
\end{figure}

Figure \ref{fig:HExt} shows the reddening distribution of our star cluster sample. The distribution predominantly peaks in the range \(E(B-V) \sim 0.13\) to \(0.15\) mag, in good agreement with findings reported in the literature \citep[see, e.g.,][and references therein]{2007ApJ...662..969I}.

\begin{figure}
    \begin{center}
        \includegraphics[width=1.0\columnwidth]{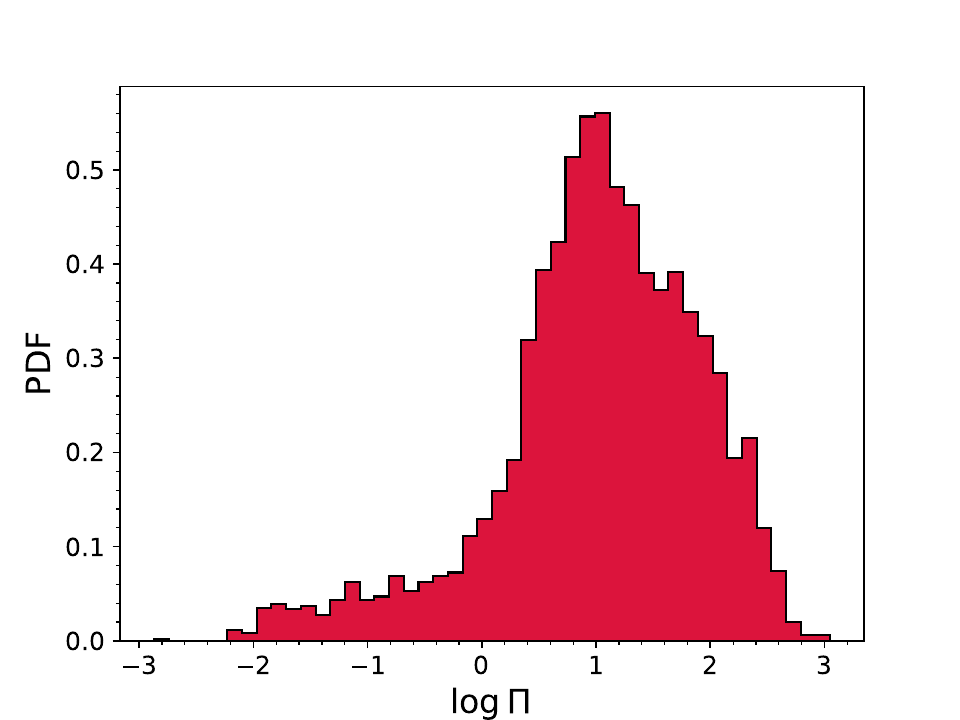}
    \end{center}
	\caption{Distribution of boundedness parameter $\Pi$ for our sample.}
    \label{fig:Hpi}
\end{figure}

In Figure \ref{fig:Hpi}, we show the results for the boundedness parameter $\Pi$ as defined by \citet{2011MNRAS.410L...6G}. By their definition, all objects with $\Pi > 1$ are bound, so it is clear that most of our objects are indeed proper star clusters.

\begin{figure*}
\begin{center}
        \includegraphics[width=1.65\columnwidth]{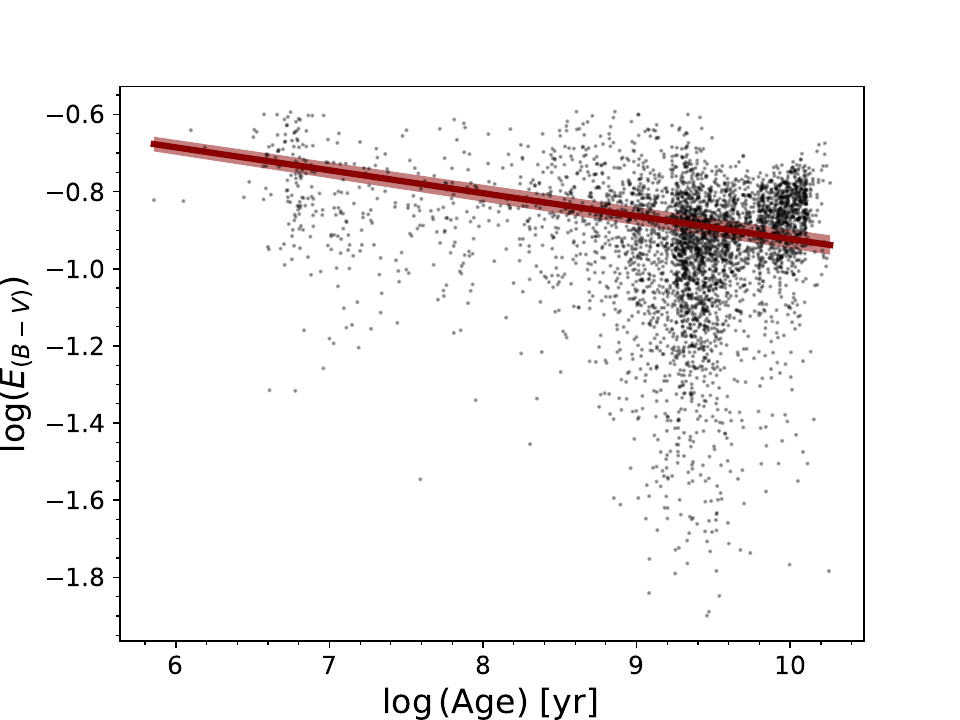}
\end{center}
	\caption{Age-extinction relation in our sample. {\it Dark red line:} best fit; {\it shaded red region:} 1$\sigma$ uncertainties.}
\label{fig:AErel}
\end{figure*}

In Figure \ref{fig:AErel}, we show the age-extinction relation for our star cluster sample. Here, we executed a linear regression analysis on the data, accounting for error bars along both axes, employing an MCMC approach. We clearly see an inverse correlation between age and extinction: younger clusters have larger extinction than their older counterparts. This result is consistent with the literature \citep[e.g.,][]{char00}.

\begin{figure*}
\begin{center}
        \includegraphics[width=1.5\columnwidth]{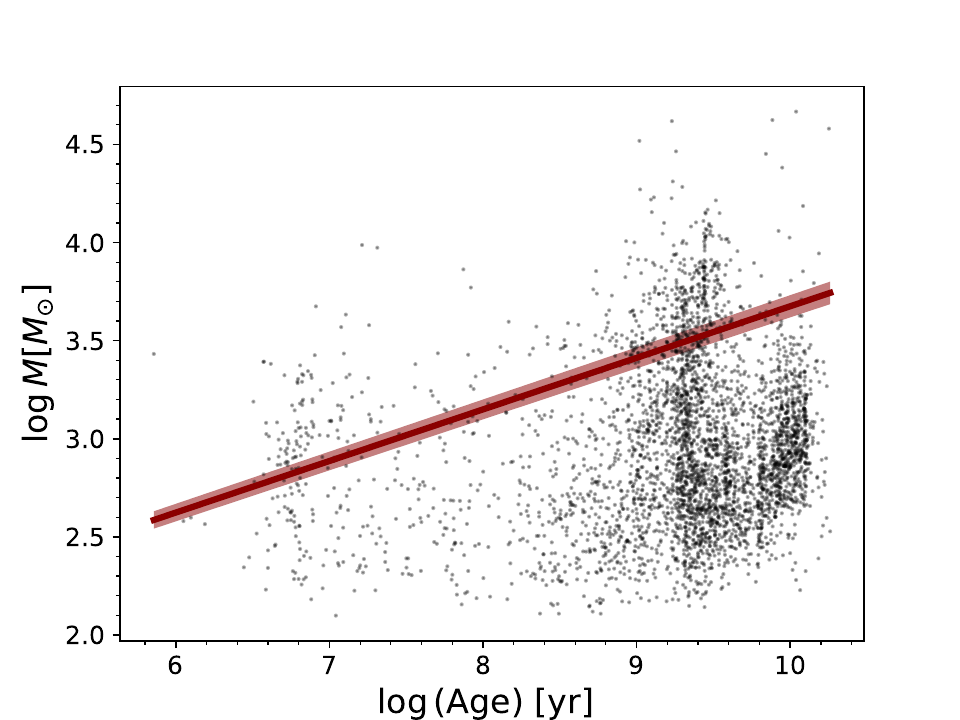}
\end{center}
	\caption{Mass-Age relation in our sample. {\it Dark red line:} best fit; {\it shaded red region:} 1$\sigma$ uncertainties.}
\label{fig:MArel}
\end{figure*}

In Figure \ref{fig:MArel}, we display the observed age-mass relation for our sample of star clusters. Here, we rigorously account for error bars in both the age and mass dimensions, utilising an MCMC approach to ensure robust statistical inference. Our analysis reveals a discernible correlation between the age and mass of star clusters: notably, younger clusters exhibit a tendency towards lower mass compared to their older counterparts. This observation aligns with existing findings in astrophysical research, suggesting a consistent pattern in the age-mass relationship of star clusters \citep[see][]{de2006well, 2008A&A...482..165G}.

\subsection{Comparison with other catalogues}
\begin{figure}
\begin{center}
        \includegraphics[width=1.\columnwidth]{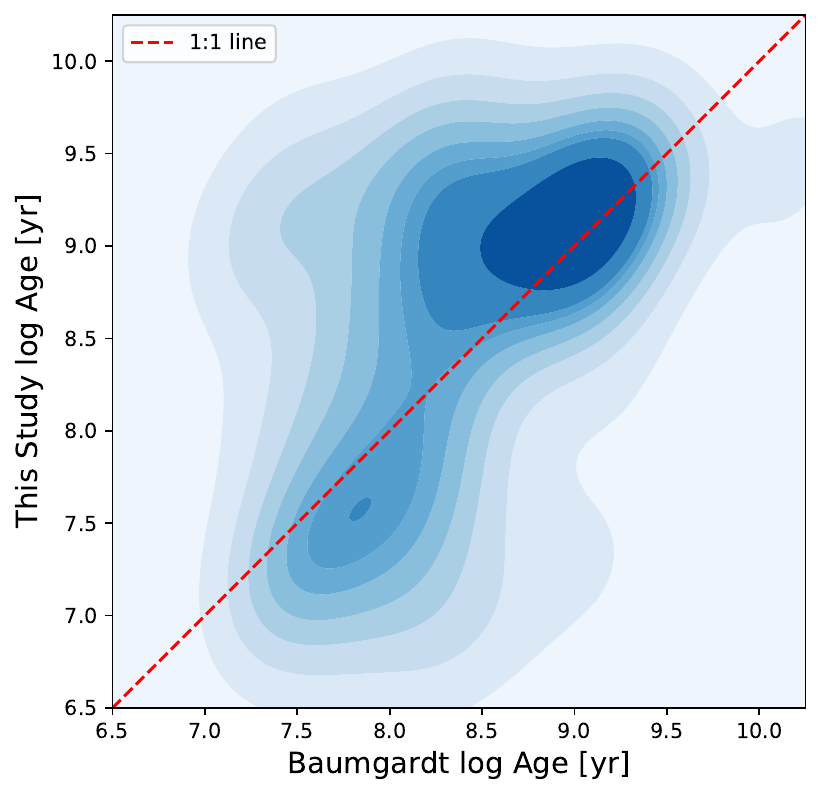}
\end{center}
	\caption{Comparison of our estimated cluster ages with those from \citet{2013MNRAS.430..676B}. The density distribution of age estimates is shown as filled contours, where darker shades indicate higher concentrations of data points. The red dashed line represents the 1:1 relation, corresponding to perfect agreement between the two datasets. }
\label{fig:AgeBaumComp}
\end{figure}

In this section, we compare our estimated physical parameters with those reported in previously published catalogues. Specifically, we examine the overlap with the catalogue presented by \citet{2013MNRAS.430..676B}, and identify 244 common objects out of their total sample of 320. Furthermore, in Figure \ref{fig:AgeBaumComp}, we present a direct comparison of our age estimates of the clusters with those obtained by \citet{2013MNRAS.430..676B}; while a general correlation is observed, deviations from the 1:1 line highlight systematic differences in age determinations, potentially due to methodological differences or varying input data.

Compared with the catalogue presented in \citet{Bica2008}, we find 1132 common objects out of their total sample of 3740,  while in comparison  with the catalogue presented in \citet{Bitsakis2017}, we have 3962 objects in common out of their total sample of 4850.

\begin{figure}
\begin{center}
        \includegraphics[width=1.\columnwidth]{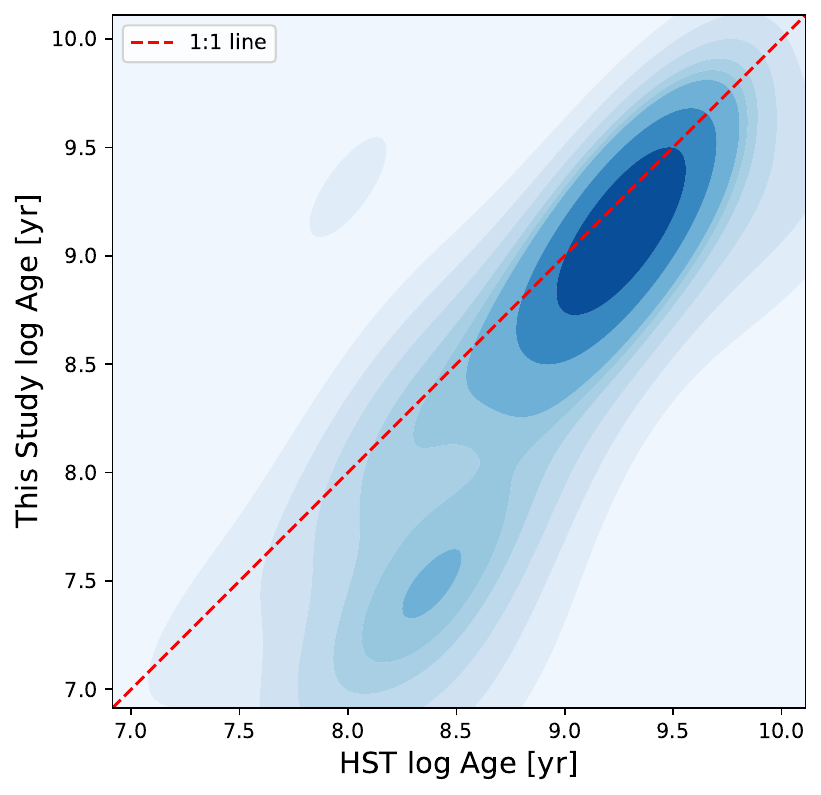}
\end{center}
	\caption{Comparison of our estimated cluster ages with those from \citet{2023A&A...672A.161M}. The density distribution of age estimates is shown as filled contours, where darker shades indicate higher concentrations of data points. The red dashed line represents the 1:1 relation, corresponding to perfect agreement between the two datasets. }
\label{fig:HSTagecomp}
\end{figure}

We also compare our results with star clusters detected using the \textit{Hubble Space Telescope} (HST) data, as published by \citet{2023A&A...672A.161M}, and identify 30 clusters in common. Figure \ref{fig:HSTagecomp} presents a comparison between our age estimates and those from \citet{2023A&A...672A.161M}, showing a general agreement within the uncertainties, with both datasets following the same overall trend.

Additionally, we conducted our full ASteCA analysis using the HST photometry published by \citet{2023A&A...672A.161M} for four clusters: KMHK~1231, NGC~1751, NGC~1818, and NGC~1846. Expanding the analysis to additional clusters was not feasible,  due to either an insufficient number of available photometric bands or the absence of reported uncertainties for one or more of them. The results of this analysis are fully consistent with those obtained using SMASH photometry, with all values agreeing within the uncertainties.

\section{The most massive stars in the youngest star clusters}
To investigate the properties of young stellar clusters, we selected a sub-sample based on the criterion $\log (\mathrm{age/yr}) \leq 6.7$, corresponding to clusters younger than 5 Myr. This refinement yields 109 clusters. Using a random posterior resampling, we estimate that up to 20.6\% of this young cluster sub-sample may be contaminated by older clusters.

The \( m_{\text{max}}-M_{\text{cl}} \) relation, which correlates the mass of the most massive star in a star cluster (\( m_{\text{max}} \)) with the total mass of the star cluster itself (\( M_{\text{cl}} \)), is a pivotal concept in understanding the influence of deterministic processes in star formation. Recent studies challenge the traditional stochastic interpretations of the IMF, suggesting instead that star formation within clusters is more self-regulated and dependent on environmental factors \citep{weidner2006, 2023A&A...670A.151Y}. 

However, \citet{2008MNRAS.391..711M} have revisited the question of whether the maximum stellar masses within star clusters can be explained through stochastic sampling from an IMF, in turn challenging assertions of deterministic processes like optimal sampling. They compiled a substantial dataset of maximum stellar mass versus cluster member numbers from various observational studies, and compared it with the predictions of a stochastic model using statistical tools. Their approach involved reassessing how selection methods and cluster membership assignments impact the observed maximum stellar masses. Despite previous studies suggesting that lower maximum stellar masses in smaller clusters may indicate a deviation from stochastic behavior, \citet{2008MNRAS.391..711M} found that the data largely support the random drawing hypothesis,
and that any perceived deviations are within the expected variability.
This conclusion disputes the notion that the mass of the most massive star is strictly determined by cluster properties, suggesting instead that the broad statistical distributions observed can be attributed to the inherent randomness in the sampling of stellar masses from a universal IMF.

In another study, \citet{Lamb_2010} investigated the formation environments of massive stars in the Small Magellanic Cloud, leveraging observations with the Hubble Space Telescope. They categorised eight OB stars as either truly isolated or part of sparse clusters with few companion stars. Employing Monte Carlo simulations, this paper evaluated whether a universal IMF can explain these minimal clusters, by testing both core accretion and  competitive accretion theories of star formation. The findings of \citet{Lamb_2010} align better with the core accretion model, suggesting that these stars can arise in a less dense environment without a
strong correlation with cluster mass; thus, they support the stochastic formation of massive stars in sparsely populated clusters. And in yet another work, from radiation magnetohydrodynamic  simulations of low-mass star cluster formation, \citet{2023OJAp....6E..48G} conclude that stellar masses  are not random and do not follow stochastic sampling while a cluster is forming, although stochastic sampling may describe the end result well. Hence the importance of studying very young clusters.

Stochastic sampling treats star formation as a series of independent random events, in which the mass of each star is drawn randomly from an invariant probability density function, typically the IMF. According to stochastic sampling, there is significant variability expected in the mass of the most massive star, in relation to the total cluster mass. This hypothesis predicts a wider dispersion in the \( m_{\text{max}}-M_{\text{cl}} \) relation, and a high degree of randomness in the formation and distribution of stellar masses within clusters. In contrast, optimal sampling is proposed as a deterministic phenomenon, according to which the formation of all stars in a cluster, especially the massive ones, is highly correlated, due to self-regulation mechanisms within the star formation process. This assumption implies a fixed relationship between the mass of a star cluster and the mass of the most massive star it contains. The optimal sampling model results in a much tighter \( m_{\text{max}}-M_{\text{cl}} \) distribution, and is consistent with observations that show less scatter than what stochastic models predict.

Observational evidence points to a tighter correlation in the \( m_{\text{max}}-M_{\text{cl}} \) relation than would be expected from purely random sampling of stellar masses, indicating that the cluster's mass has a direct influence on the maximum stellar mass it can produce \citep{2023A&A...670A.151Y, 2015MNRAS.449.1327V}. This view is supported by theoretical models that propose mechanisms, such as competitive accretion and feedback processes, which regulate star formation and potentially limit the mass of stars within less massive clusters \citep{bonnell2003, 2006MNRAS.370..488B}. Consequently, the \( m_{\text{max}}-M_{\text{cl}} \) relation not only provides insights into the efficiency of star formation across different environments, but also impacts the broader understanding of galactic evolution and dynamics within stellar nurseries.

To test the statistical and optimal sampling hypotheses, we estimate cluster masses with the ASteCA software suite and its isochrone fitting methodology. As described in detail in Section \S 4, this method involves generating synthetic clusters from theoretical isochrones, which are then used to find the best match with observed cluster data. The optimisation of this match is achieved through an MCMC genetic algorithm. This approach not only facilitates a robust and accurate determination of stellar cluster masses, but also quantifies their associated uncertainties.

To estimate the masses of individual stars within the clusters, we use their optical photometric data, again processed through the ASteCA software suite. As already mentioned, ASteCA incorporates a Bayesian field-star decontamination algorithm that assigns membership probabilities to stars, based on their photometry. Upon establishing these probabilities, we generate a list of confirmed cluster members, from which the most massive star within the cluster can be identified.

\begin{figure*}
\begin{center}
        \includegraphics[width=1.5\columnwidth]{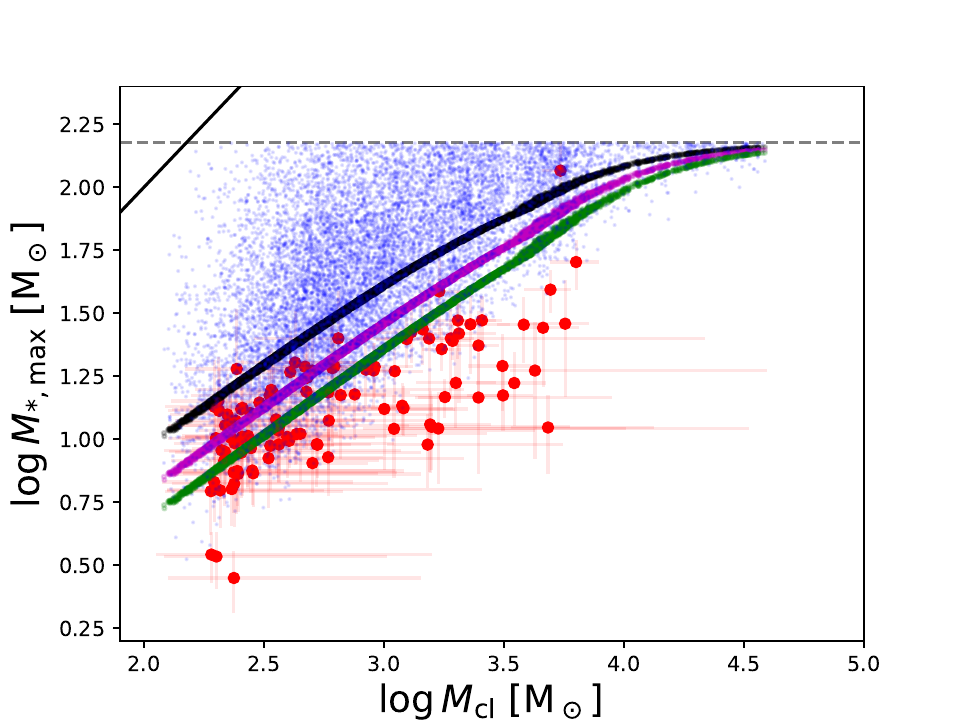}
\end{center}
	\caption{Most massive star mass to young cluster mass relation. The optimally sampled results for the first, second, and third most massive star as a function of embedded cluster mass are shown as a black, a magenta,  and a green thick solid curve, respectively; the thicknesses of the lines are given by random realisations inside the $M_{\mathrm{cl}}$ errors. The blue cloud is the randomly sampled most massive star given by random realisations inside the $M_{\mathrm{cl}}$ errors. The thin solid line indicates the $m_{\mathrm{max}} = M_{\mathrm{cl}}$ limit; the horizontal thin dashed line marks the 150 M$_\odot$ limit.}
\label{fig:MsMrel}
\end{figure*}

For the youngest star clusters sub-sample, we probe the nature of star formation by examining the $m_{\mathrm{max}}-M_{\mathrm{cl}}$ relation, shown in Figure \ref{fig:MsMrel}. 
The figure displays the optimally sampled results for the first, second, and third most massive stars, as a function of embedded cluster mass. They are shown as a black, a magenta, and a green thick solid curve, respectively. The thicknesses of the lines are given by uniformly sampled realisations inside the $M_{\mathrm{cl}}$ uncertainty distributions. The blue cloud is the randomly sampled most massive star, given by stochastic realisations inside the embedded cluster mass uncertainties.
We can see that our youngest stellar clusters sub-sample seems to follow, within the uncertainties, the optimal sampling tendency, and with less dispersion than the randomly sampled cloud. This seems to align with the recent findings in the Milky Way obtained by \citet {2023A&A...670A.151Y}. Our results are, in general, consistent with their optimally sampled model.

Ultimately, the debate between optimal and stochastic sampling carries significant implications for our understanding of galaxy evolution, the feedback from massive stars, and the interpretation of integrated light from unresolved stellar populations. While our work aims to contribute to this ongoing discussion by presenting and analysing data from the LMC, it represents only a modest step towards resolving these complex issues. Further observational efforts and the development of improved theoretical models are essential to reconcile the differences between these sampling scenarios, and to achieve a more unified understanding of how stars populate the IMF. We view our contribution as part of a broader, collaborative effort that highlights the need for continued research in this area.

\section{Conclusions}
In this paper, we presented a detailed study of the youngest star clusters in the LMC. Our research primarily focused on understanding the physical properties, scaling laws, and stellar IMF of these clusters. 

In the study, the multi-wavelength data approach significantly enhanced the analysis of the youngest star clusters in the LMC. This comprehensive methodology incorporated observations from infrared (SAGE), optical (SMASH), and ultraviolet (GALEX and SUMAC) surveys. By using these diverse data sources, we were able to create a more detailed and nuanced understanding of the physical properties of these clusters, such as age, mass, and metallicity. This holistic view is critical for accurately assessing the characteristics and evolutionary stages of these young clusters.

The adoption of the BIRCH  algorithm marked a significant advancement in cluster detection methodology. This approach was particularly effective in parsing the extensive multi-wavelength data, allowing for the precise identification and classification of the youngest star clusters in the LMC. The BIRCH algorithm's ability to efficiently handle large datasets and distinguish subtle differences in cluster characteristics contributed to a more accurate and detailed understanding of cluster distribution, sizes, and ages within the LMC. This innovative technique proved to be a key factor in the success of the study, demonstrating its potential for broader applications in astronomical research.

The Markov Chain Monte Carlo analysis, conducted using the ASteCA code, played a pivotal role in uncovering the key physical parameters of the youngest star clusters in the LMC. This advanced statistical technique allowed for a detailed exploration of the clusters' ages, masses, metallicities, and extinctions. The implementation of ASteCA enabled a more refined and accurate determination of these parameters. This can potentially lead to an enhanced understanding of the clusters' intrinsic properties, and hence to a better knowledge of the star formation processes in the LMC.

This study's exploration of the youngest star clusters in the LMC provided valuable insights into the early stages of stellar and galactic development.
In particular, we found a positive correlation between cluster mass and maximum stellar mass in the clusters younger than 5 Myr, confirming a recent result in the Milky Way \citep{2023A&A...670A.151Y}. These investigations may contribute to  
a deeper understanding of the mechanisms and conditions that influence star formation and cluster evolution, 
as well as of how dwarf galaxies like the LMC fit in the history of cosmic star formation and galactic evolution.

\section*{Acknowledgements}
R.C. acknowledges the financial support from the CONAHCYT research grant CF2022-320152. R.A.G.L. acknowledges the financial support of DGAPA, UNAM, projects IN108518 and IN106124, and PASPA program, and of CONAHCYT, Mexico, project A1-S-8263. GB acknowledges financial support from the National Autonomous University of M\'exico (UNAM) through grants DGAPA/PAPIIT IG100319 and BG100622.

\section*{Data availability}
The datasets supporting the conclusions of this article, including the data used for generating the figures, are available from the corresponding author upon reasonable request.

\bibliography{bib/bib2022}

\label{lastpage}

\appendix
\section{Catalogue}
In Table \ref{tab:tab01} we provide a detailed catalogue of the 3801 star clusters detected in the LMC, as described in section \S~3, with their physical properties obtained as described in section \S~4. The table is structured to facilitate easy access to the data for each cluster, and comprises the following columns:

\begin{itemize}
    \item Column (1): \textbf{Cluster Index} -- A unique numerical identifier for each cluster.
    \item Column (2): \textbf{Cluster Name} -- The designation of the cluster.
    \item Column (3): \textbf{Right Ascension (degrees)} -- The right ascension of the cluster, expressed in degrees.
    \item Column (4): \textbf{Declination (degrees)} -- The declination of the cluster, in degrees.
    \item Column (5): \textbf{Radius (parsecs)} -- The approximate size of the cluster from its center to its outer edges, in parsecs.
    \item Column (6): \textbf{Core Radius (parsecs)} -- The radius of the cluster's denser core area, in parsecs.
    \item Column (7): \textbf{Tidal Radius (parsecs)} -- The outer boundary of the cluster where gravitational influences from external bodies become significant, in parsecs.
    \item Column (8): \textbf{Log(Age [years])} -- The decimal logarithm of the cluster's age, in years.
    \item Column (9): \textbf{Internal Extinction (magnitudes)} -- The measure of light absorption and scattering due to internal dust, in magnitudes.
    \item Column (10): \textbf{Log(Mass [Solar Masses])} -- The decimal logarithm of the cluster’s mass, in solar masses.
    \item Column (11): \textbf{Cluster CIDS} -- The names given to this cluster in other works, first if present in those catalogues, as in \citet{2013MNRAS.430..676B} and \citet{Bica2008} as well as other common names and then as in \citet{Bitsakis2017}.
\end{itemize}

This catalogue serves as a valuable resource for researchers studying the properties and evolutionary stages of star clusters at the LMC. 

\onecolumn
\begin{landscape}
\scriptsize
\begin{longtable}{@{} c l r r r r r r r r c l @{}}

\caption{Star cluster catalogue. Column (1) is the cluster's index, column (2) is the cluster's name, column (3) is the cluster's right ascension in degrees, column (4) is the cluster's declination in degrees, column (5) is the cluster's radius in parsecs,  column (6) is the core radius in parsecs, column (7) is the tidal radius in parsecs, column (8) is the decimal logarithm of the cluster's age in years, column (9) is the cluster's internal extinction in magnitudes, column (10) is the decimal logarithm of the cluster's  mass in solar masses, and column (11) other common names to the cluster.}
\label{tab:tab01} \\

\hline \hline \\[-2ex]
\multicolumn{1}{c}{(1)} &
\multicolumn{1}{c}{(2)} &
\multicolumn{1}{c}{(3)} &
\multicolumn{1}{c}{(4)} &
\multicolumn{1}{c}{(5)} &
\multicolumn{1}{c}{(6)} &
\multicolumn{1}{c}{(7)} &
\multicolumn{1}{c}{(8)} &
\multicolumn{1}{c}{(9)} &
\multicolumn{1}{c}{(10)} &
\multicolumn{1}{c}{(11)} \\

\multicolumn{1}{c}{\bf{Index}} &
\multicolumn{1}{c}{\bf{Name}} &
\multicolumn{1}{c}{$\alpha$(J2000) } &
\multicolumn{1}{c}{$\delta$(J2000) } &
\multicolumn{1}{c}{$\mathbf{R_{cl}}$} &
\multicolumn{1}{c}{$\mathbf{R_{c}}$} &
\multicolumn{1}{c}{$\mathbf{R_{t}}$} &
\multicolumn{1}{c}{$\mathbf{\log a}$} &
\multicolumn{1}{c}{$\mathbf{E_{(B-V)}}$} &
\multicolumn{1}{c}{$\mathbf{\log M}$} &
\multicolumn{1}{c}{\bf{CIDS}}\\

\multicolumn{1}{c}{} &
\multicolumn{1}{c}{} &
\multicolumn{1}{c}{(deg)} &
\multicolumn{1}{c}{(deg)} &
\multicolumn{1}{c}{(pc)} &
\multicolumn{1}{c}{(pc)} &
\multicolumn{1}{c}{(pc)} &
\multicolumn{1}{c}{($\mathrm{yr}$)} &
\multicolumn{1}{c}{($\mathrm{mag}$)} &
\multicolumn{1}{c}{($\mathrm{M_{\odot}}$)} &
\multicolumn{1}{c}{}\\[0.5ex] \hline \\[-1.8ex]
\endfirsthead

\multicolumn{1}{c}{(1)} &
\multicolumn{1}{c}{(2)} &
\multicolumn{1}{c}{(3)} &
\multicolumn{1}{c}{(4)} &
\multicolumn{1}{c}{(5)} &
\multicolumn{1}{c}{(6)} &
\multicolumn{1}{c}{(7)} &
\multicolumn{1}{c}{(8)} &
\multicolumn{1}{c}{(9)} &
\multicolumn{1}{c}{(10)} &
\multicolumn{1}{c}{(11)} \\

\multicolumn{1}{c}{\bf{Index}} &
\multicolumn{1}{c}{\bf{Name}} &
\multicolumn{1}{c}{$\mathbf{\alpha}$(J2000) } &
\multicolumn{1}{c}{$\mathbf{\delta }$ (J2000) } &
\multicolumn{1}{c}{$\mathbf{R_{cl}}$} &
\multicolumn{1}{c}{$\mathbf{R_{c}}$} &
\multicolumn{1}{c}{$\mathbf{R_{t}}$} &
\multicolumn{1}{c}{$\mathbf{\log a}$} &
\multicolumn{1}{c}{$\mathbf{E_{(B-V)}}$} &
\multicolumn{1}{c}{$\mathbf{\log M}$} &
\multicolumn{1}{c}{\bf{CIDS}}\\

\multicolumn{1}{c}{} &
\multicolumn{1}{c}{} &
\multicolumn{1}{c}{(deg)} &
\multicolumn{1}{c}{(deg)} &
\multicolumn{1}{c}{(pc)} &
\multicolumn{1}{c}{(pc)} &
\multicolumn{1}{c}{(pc)} &
\multicolumn{1}{c}{($\mathrm{yr}$)} &
\multicolumn{1}{c}{($\mathrm{mag}$)} &
\multicolumn{1}{c}{($\mathrm{M_{\odot}}$)} &
\multicolumn{1}{c}{} \\[0.5ex] \hline \\[-1.8ex]

\endhead

\multicolumn{2}{l}{{Continued on Next Page\ldots}} \\

\endfoot

\\[-1.8ex] \hline 

\endlastfoot

0 & J043554.94-694545.72 & 68.9789 & -69.7627 & 6.78 & 6.22 & 10.96 & 9.45 $\pm$ 1.36 & 0.14 $\pm$ 0.07 & 2.70 $\pm$ 1.05 & \verb|NUV_1313| \\
1 & J043642.34-695132.76 & 69.1764 & -69.8591 & 8.56 & 6.17 & 13.06 & 9.44 $\pm$ 0.86 & 0.06 $\pm$ 0.06 & 2.99 $\pm$ 0.63 & \verb|IR1_1263| \\
2 & J043721.26-691209.36 & 69.3386 & -69.2026 & 40.05 & 33.06 & 39.66 & 9.30 $\pm$ 0.26 & 0.07 $\pm$ 0.03 & 3.60 $\pm$ 0.10 & \verb|NUV_1387| \\
3 & J043732.59-703501.32 & 69.3858 & -70.5837 & 50.75 & 33.94 & 38.51 & 9.30 $\pm$ 0.02 & 0.09 $\pm$ 0.01 & 4.28 $\pm$ 0.06 & \verb|NGC1651,SL7,LW12,ESO55SC30,IR1_1231| \\
4 & J043749.92-690145.12 & 69.458 & -69.0292 & 68.15 & 21.23 & 25.98 & 9.35 $\pm$ 0.04 & 0.05 $\pm$ 0.01 & 4.08 $\pm$ 0.14 & \verb|SL8,LW13,KMHK21IR1_1308| \\
5 & J043822.54-684021.00 & 69.5939 & -68.6725 & 13.06 & 15.63 & 20.52 & 9.30 $\pm$ 0.38 & 0.06 $\pm$ 0.06 & 3.33 $\pm$ 0.35 & \verb|NGC1652,SL10,LW14,ESO55SC32,IR1_1351| \\
6 & J043833.34-693126.04 & 69.6389 & -69.5239 & 10.93 & 10.83 & 18.04 & 9.49 $\pm$ 1.48 & 0.11 $\pm$ 0.07 & 2.78 $\pm$ 1.03 & \verb|NUV_1368| \\
7 & J043908.62-694610.92 & 69.7859 & -69.7697 & 23.00 & 31.76 & 41.48 & 9.36 $\pm$ 0.19 & 0.03 $\pm$ 0.03 & 3.33 $\pm$ 0.16 & \verb|NUV_1335| \\
8 & J043915.00-692914.28 & 69.8125 & -69.4873 & 23.25 & 15.12 & 23.07 & 9.44 $\pm$ 0.41 & 0.05 $\pm$ 0.03 & 3.61 $\pm$ 0.12 & \verb|NUV_1376| \\
9 & J043953.74-685440.68 & 69.9739 & -68.9113 & 5.65 & 6.00 & 9.78 & 10.10 $\pm$ 1.14 & 0.15 $\pm$ 0.07 & 2.73 $\pm$ 4.44 & \verb|NUV_1454| \\
10 & J043959.28-691624.96 & 69.997 & -69.2736 & 16.22 & 17.08 & 25.87 & 9.46 $\pm$ 0.86 & 0.12 $\pm$ 0.08 & 2.92 $\pm$ 0.85 & \verb|NUV_1404| \\
\dots & \dots 			&	 \dots  & 		\dots & \dots & \dots &\dots & \dots  			& \dots 			& \dots		      &			 \dots  \\
3952 & J060303.48-672358.20 & 90.7645 & -67.3995 & 8.90 & 9.90 & 15.62 & 9.04 $\pm$ 0.60 & 0.16 $\pm$ 0.08 & 2.74 $\pm$ 1.36 & \verb|LW368,BM214,KMHK1610IR1_2239| \\
3953 & J060328.58-670623.04 & 90.8691 & -67.1064 & 16.06 & 19.89 & 27.51 & 9.29 $\pm$ 0.72 & 0.14 $\pm$ 0.07 & 2.59 $\pm$ 1.07 & \verb|NUV_1246| \\
3954 & J060341.45-670907.92 & 90.9227 & -67.1522 & 54.71 & 22.66 & 32.77 & 9.25 $\pm$ 0.03 & 0.02 $\pm$ 0.01 & 3.74 $\pm$ 0.08 & \verb|NUV_1212| \\
3955 & J060354.98-675633.72 & 90.9791 & -67.9427 & 3.89 & 3.50 & 6.45 & 9.08 $\pm$ 0.76 & 0.14 $\pm$ 0.07 & 2.38 $\pm$ 2.11 & \verb|NUV_1088| \\
3956 & J060409.58-672636.60 & 91.0399 & -67.4435 & 3.84 & 3.32 & 5.99 & 9.54 $\pm$ 0.71 & 0.08 $\pm$ 0.07 & 2.72 $\pm$ 0.77 & \verb|NUV_1160| \\
3957 & J060432.16-675448.96 & 91.134 & -67.9136 & 4.00 & 2.08 & 5.10 & 9.95 $\pm$ 0.92 & 0.15 $\pm$ 0.07 & 3.04 $\pm$ 0.64 & \verb|NUV_1092| \\
3958 & J060436.43-673115.24 & 91.1518 & -67.5209 & 21.32 & 20.24 & 28.17 & 9.04 $\pm$ 0.14 & 0.04 $\pm$ 0.02 & 2.86 $\pm$ 0.22 & \verb|NUV_1148| \\
3959 & J060447.18-682201.56 & 91.1966 & -68.3671 & 25.13 & 15.31 & 23.53 & 9.28 $\pm$ 0.55 & 0.08 $\pm$ 0.08 & 3.18 $\pm$ 0.39 & \verb|BM226NUV_1023| \\
3960 & J060503.74-674218.72 & 91.2656 & -67.7052 & 20.76 & 21.96 & 35.17 & 9.24 $\pm$ 0.33 & 0.07 $\pm$ 0.10 & 2.82 $\pm$ 0.22 & \verb|SL833,LW378,BM228,KMHK1627NUV_1117| \\
3961 & J060542.36-680933.12 & 91.4265 & -68.1592 & 4.26 & 3.81 & 6.82 & 8.89 $\pm$ 0.98 & 0.11 $\pm$ 0.07 & 2.48 $\pm$ 2.65 & \verb|LW381,BM230,KMHK1630NUV_1057| \\

\end{longtable}
\normalsize 
\end{landscape}
\twocolumn

\end{document}